\documentclass{iau} 

\usepackage{graphicx}
\usepackage{natbib}
\usepackage{hyperref}
\usepackage{amsmath}
\usepackage{newtxtext, newtxmath}
\usepackage{color}

\title[The radio environment of HD189733b]{Tuning in to the radio environment of HD189733b}

\author[R.~D.~Kavanagh et al.]
{R.~D.~Kavanagh$^{1}$
A.~A.~Vidotto$^{1}$, 
D.~\'{O}~Fionnag\'{a}in$^{1}$,
V.~Bourrier$^{2}$,
R.~Fares$^{3,4}$,
M.~Jardine$^{5}$,
Ch.~Helling$^{6}$,
C.~Moutou$^{7}$,
J.~Llama$^{8}$,
P.~J.~Wheatley$^{9}$}

\affiliation{
$^{1}$School of Physics, Trinity College Dublin, The University of Dublin, Dublin 2, Ireland \\
email: {\tt kavanar5@tcd.ie} \\
$^{2}$Observatoire de l'Universit\'e de Gen\'eve, Chemin des Maillettes 51, Versoix, CH-1290, Switzerland \\
$^{3}$Physics Department, United Arab Emirates University, P.O.~Box 15551, Al-Ain, United Arab Emirates \\
$^{4}$University of Southern Queensland, Centre for Astrophysics, Toowoomba, Queensland, 4350, Australia \\
$^{5}$SUPA, School of Physics and Astronomy, University of St Andrews, North Haugh, St Andrews, Fife, Scotland, KY16 9SS \\
$^{6}$Centre for Exoplanet Science, University of St Andrews, St Andrews KY16 9SS, UK \\
$^{7}$CNRS/CFHT, 65-1238 Mamalahoa Highway, Kamuela HI 96743, USA \\
$^{8}$Lowell Observatory, 1400 W.~Mars Hill Rd, Flagstaff.~AZ 86001.~USA \\
$^{9}$Department of Physics, University of Warwick, Coventry CV4 7AL, UK}

\pubyear{2020}
\volume{354}
\jname{Solar and Stellar Magnetic Fields: Origins and Manifestations}
\editors{(Editors go here)}

\begin{document}
\maketitle

% ###################################################
% Abstract
% ###################################################

\begin{abstract}
The hot Jupiter HD189733b is expected to be a source of strong radio emission, due to its close proximity to its magnetically active host star. Here, we model the stellar wind of its host star, based on reconstructed surface stellar magnetic field maps. We use the local stellar wind properties at the planetary orbit obtained from our models to compute the expected radio emission from the planet. Our findings show that the planet emits with a peak flux density within the detection capabilities of LOFAR. However, due to absorption by the stellar wind itself, this emission may be attenuated significantly. We show that the best time to observe the system is when the planet is near primary transit of the host star, as the attenuation from the stellar wind is lowest in this region.
\end{abstract}

% ################################################### 
% Introduction
% ###################################################

\section{Introduction}

Close-in hot Jupiters are expected to be sources of strong auroral radio emission, analogous of what is observed for the magnetised solar system planets \citep{zarka01}. This is thought to occur due to magnetic interactions between the stellar wind of the host star and the intrinsic magnetic field of the orbiting planet. However, despite the large number of hot Jupiters detected to date, along with numerous radio surveys, no sources of exoplanetary radio emission have been detected \citep{smith09, lazio10, lecavelier13, sirothia14, ogorman18}

HD189733b is one such exoplanet that is expected to emit strong low frequency radio emission. The planet orbits its host star just 0.03~au. The host star is magnetically active, with its unsigned field strength observed to vary from 18 to 42~G over a 9 year period \citep{fares17}. Here, we model the stellar wind of the host star, and use the stellar wind properties obtained from the models to predict the flux density and frequency emitted by the planet. This emission is found to be within the detection limit of LOFAR. However, we also find that the emission may be attenuated significantly by the stellar wind of the host star. A complete description of our work is published in \citet{kavanagh19}.

% ###################################################
% Stellar wind modelling
% ###################################################

\section{Modelling the stellar wind of the host star}

To model the wind of the host star, we perform 3D magnetohydrodynamic simulations using the BATSRUS code developed by \citet{powell99}, modified by \citet{vidotto12}. We use surface stellar magnetic field maps reconstructed from observations by \citet{fares17} as boundary conditions in our simulations, at the epochs 2013~Jun/Jul, 2014~Sep, and 2015~Jul. In our models, we adopt a coronal base density of 2~MK and number density of $10^{10}$~cm$^{-3}$. These values produce a stellar wind with a mass-loss rate of $3\times10^{-12}$~$M_\odot$~yr$^{-1}$, which is within the range of inferred values for other active K-stars \citep[see][]{wood04, jardine19, rodriguez19}.

\begin{figure}
\centering
\includegraphics[width = .3\textwidth]{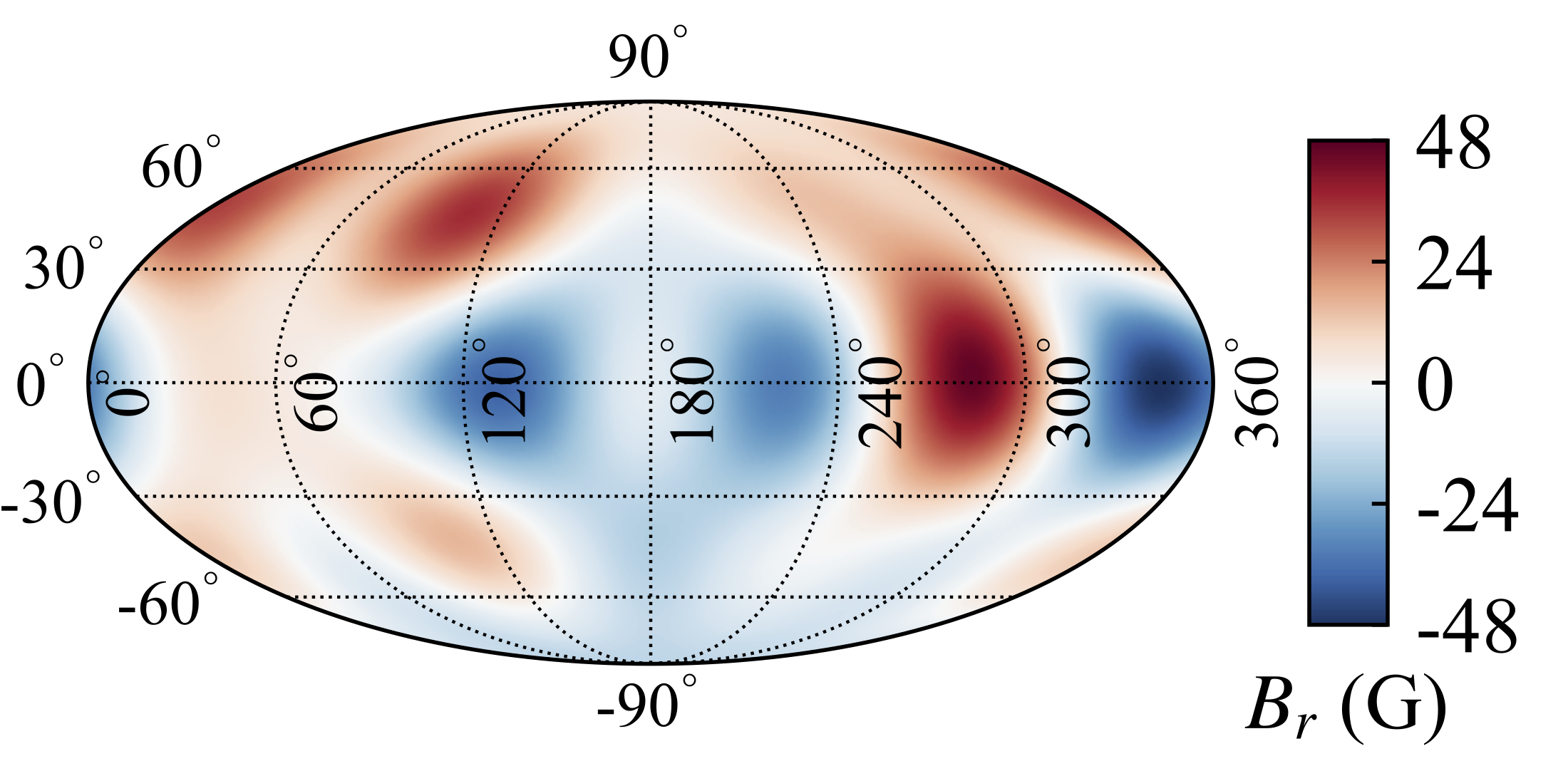}
\includegraphics[width = .3\textwidth]{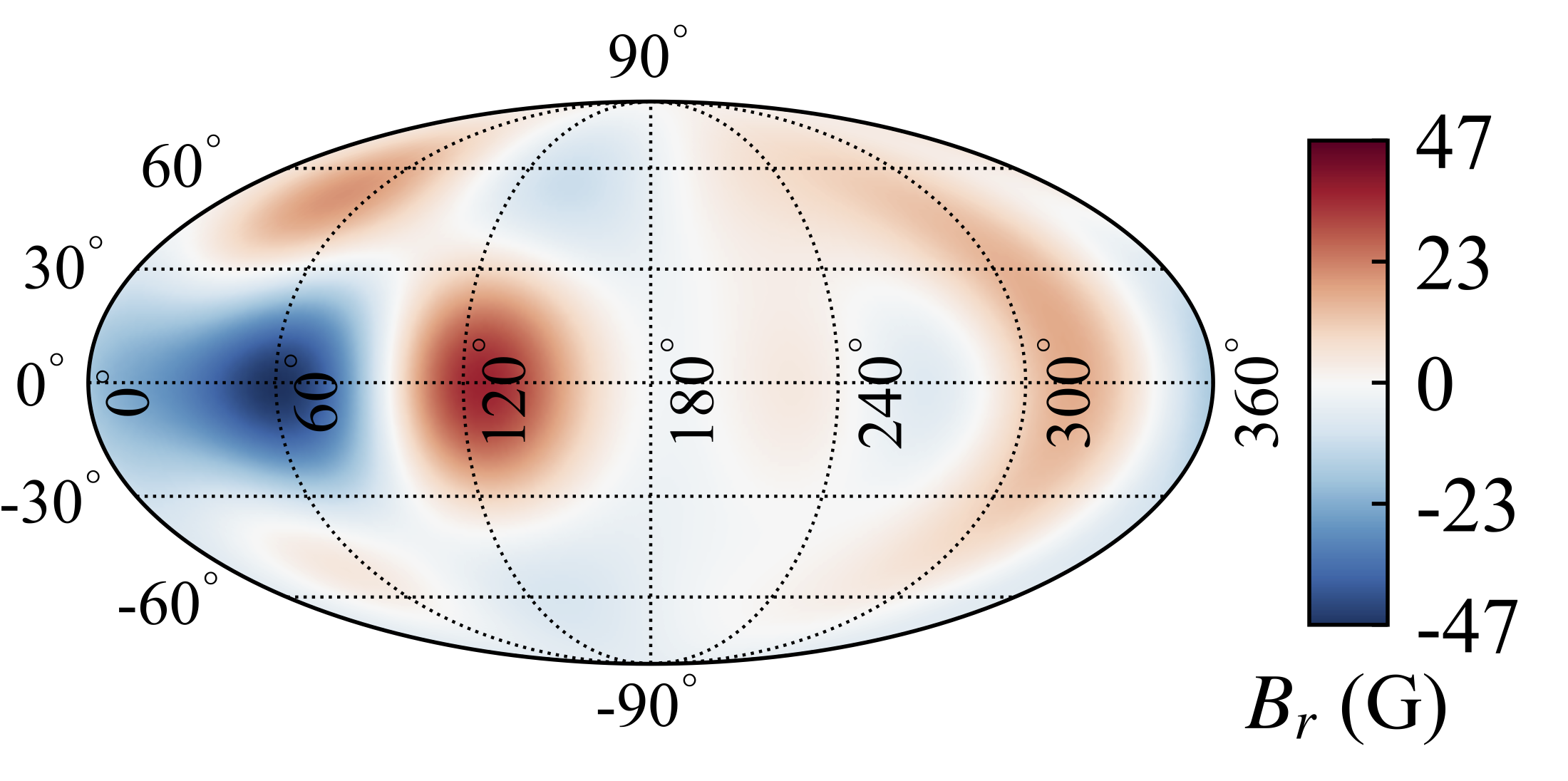}
\includegraphics[width = .3\textwidth]{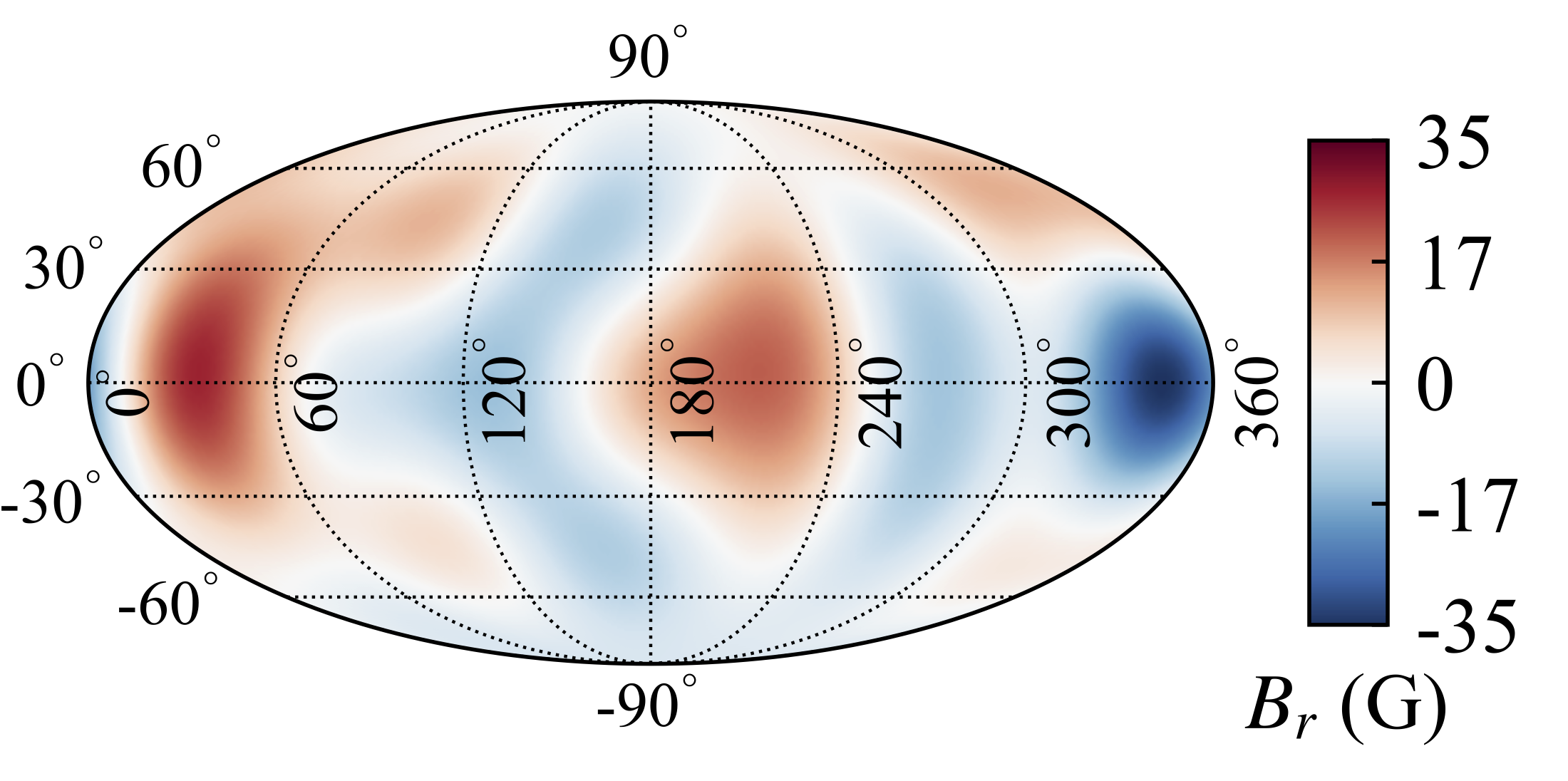}
\includegraphics[width = .3\textwidth]{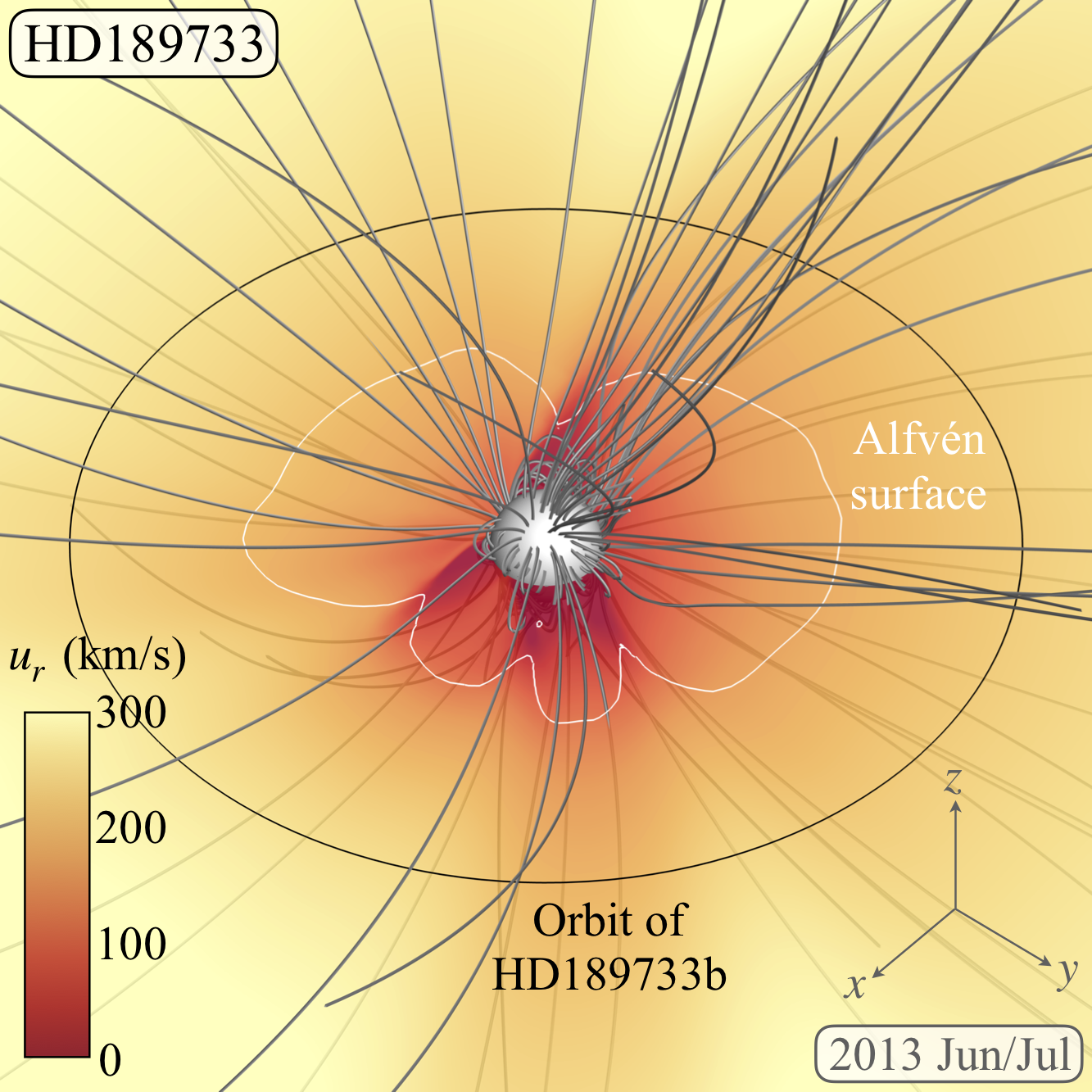}
\includegraphics[width = .3\textwidth]{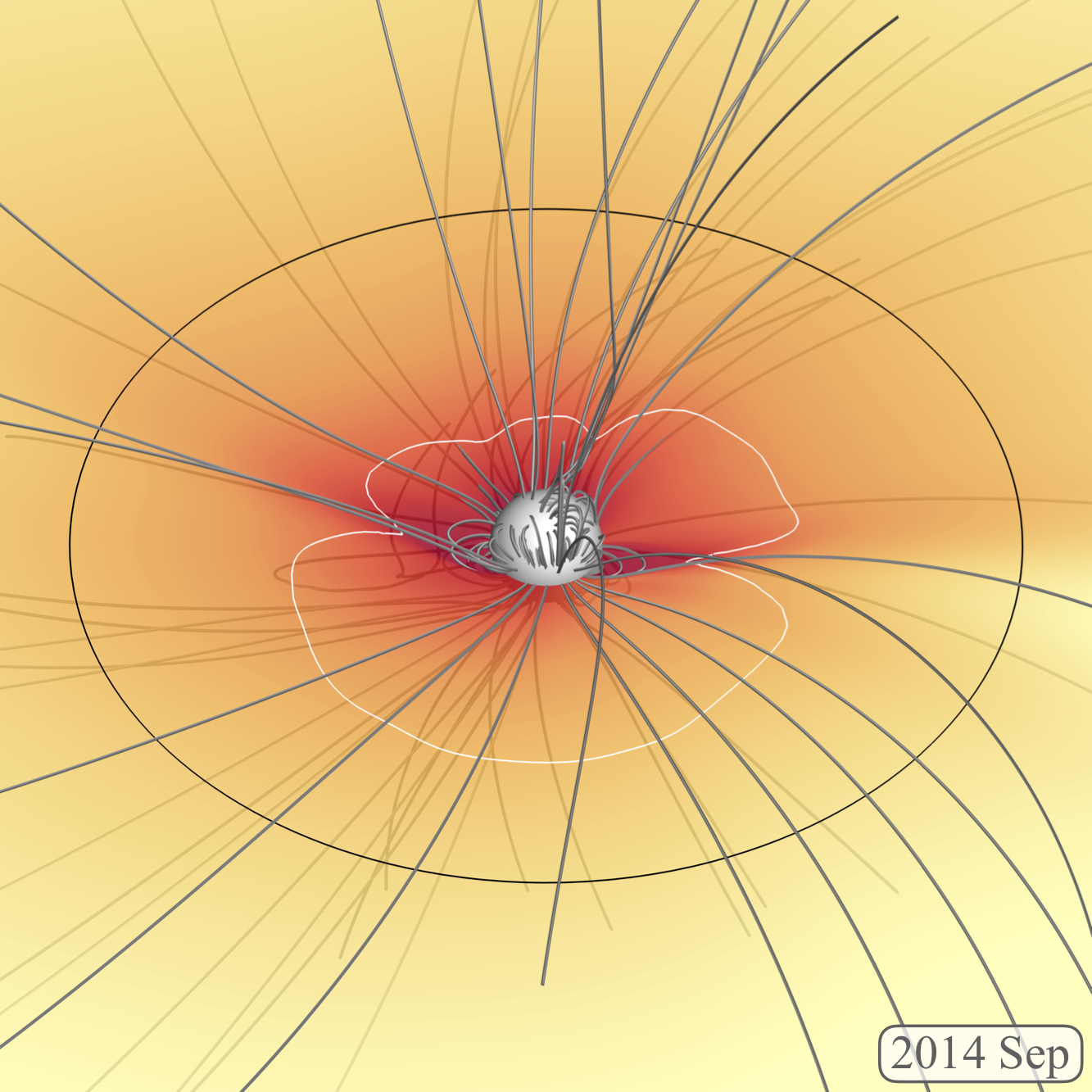}
\includegraphics[width = .3\textwidth]{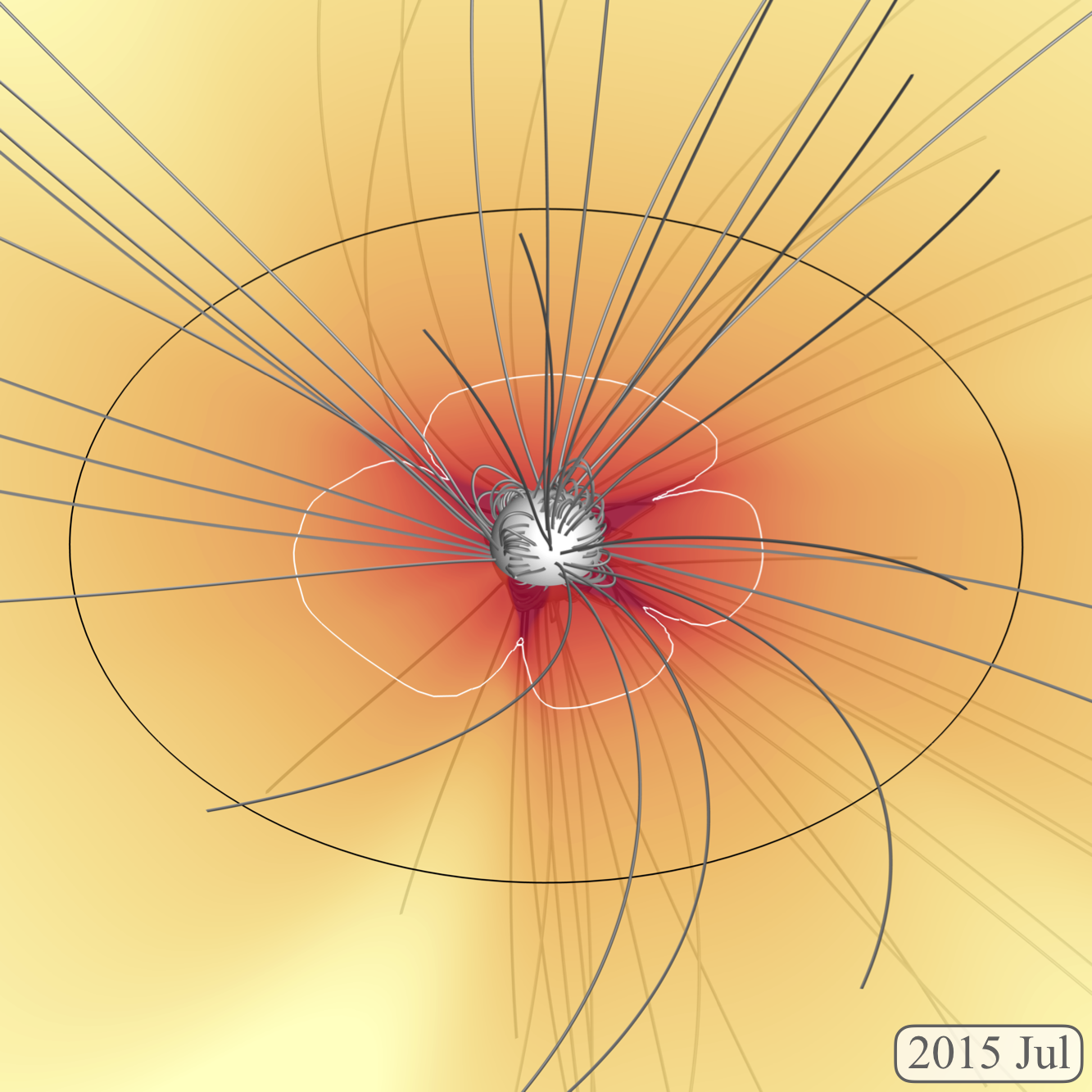}
\caption{\textit{Top panels:} Radial surface magnetic field maps of the host star reconstructed by \citet{fares17}, at the epochs 2013~Jun/Jul, 2014~Sep, and 2015~Jul (left to right). These maps are used as boundary conditions in our stellar wind simulations. \textit{Bottom panels:} Simulated stellar wind of the host star at 2013~Jun/Jul, 2014~Sep, and 2015~Jul (left to right). Grey lines show the large-scale structure of the magnetic field of the star, which is embedded in the stellar wind. Profiles of the radial velocity of the stellar wind in the orbital plane of the planet are shown. The planetary orbit is shown with a black circle, and Alfv\'en surfaces are shown in white.}
\label{fig:maps and wind}
\end{figure}

Figure~\ref{fig:maps and wind} shows the radial component of each stellar surface magnetic field map, with the corresponding wind simulation shown below each map. We see that over the modelled timescale, the stellar wind varies in response to the varying surface magnetic field. 

% ###################################################
% Planetary radio emission modelling
% ###################################################

\section{Predicting radio emission from HD189733b}

\begin{figure}
\centering
\includegraphics[width = 0.5\textwidth]{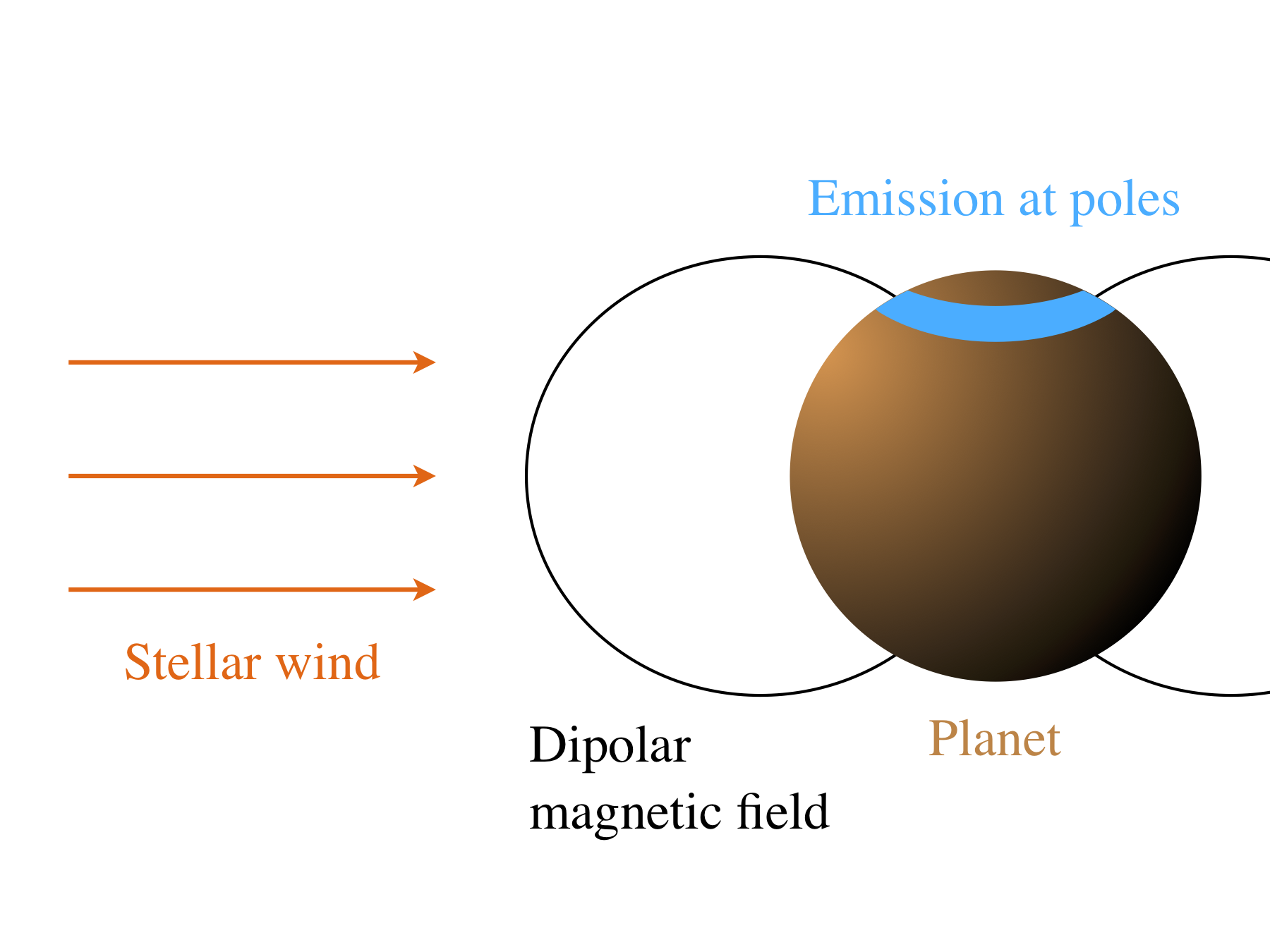}
\caption{Sketch illustrating the stellar wind incident on the magnetic field of the planet. The interaction results in radio emission from polar cap regions near the surface.}
\label{fig:sketch}
\end{figure}

Using the local stellar wind properties obtained from our simulations, we use the radiometric Bode's law to compute the expected flux density and frequency of emission from the planet, for an assumed planetary magnetic field strength \citep[see][]{vidotto17}. In the model, 0.2\% of the incident stellar wind's magnetic power is converted into radio power from the planet \citep{zarka10}. The planet's magnetic field is assumed to be dipolar. This is illustrated in Figure~\ref{fig:sketch}.

\begin{figure}
\centering
\includegraphics[width = 0.5\textwidth]{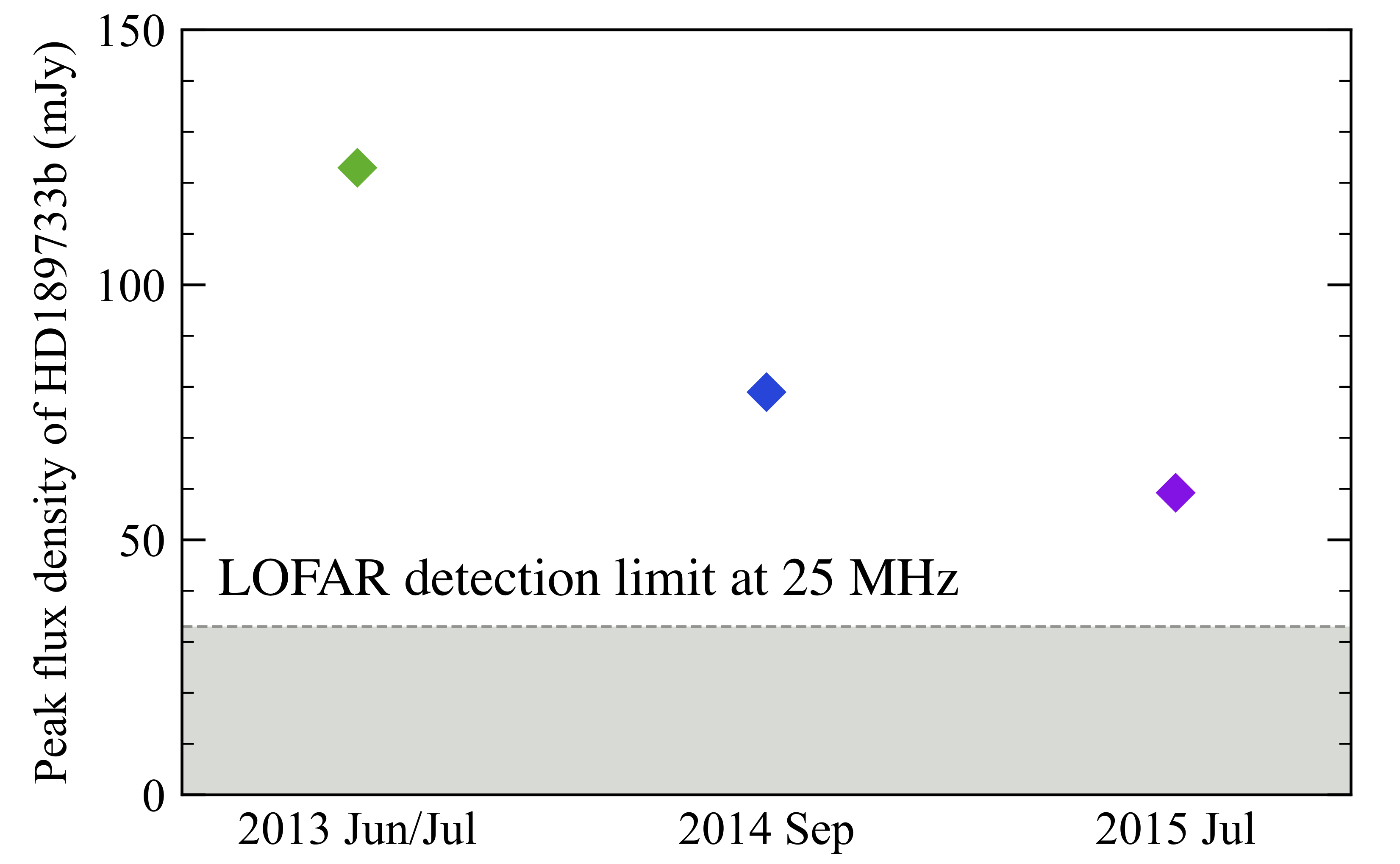}
\caption{Peak radio flux densities emitted by the planet at each modelled epoch, for a field strength of 10 G.}
\label{fig:planet emission}
\end{figure}

Figure~\ref{fig:planet emission} shows the predicted peak flux densities received at Earth from HD189733b at 2013~Jun/Jul, 2014~Sep, and 2015~Jul, computed using the stellar wind properties at the planet's orbit obtained from our models. For an assumed planetary magnetic field strength of 10~G, we find that this emission occurs at a frequency of 25~MHz. Due to the variability of the stellar wind over the three modelled epochs, the peak flux densities from the planet also vary. At 25~MHz, the emission predicted from HD189733b place it above the detection limit of LOFAR for a 1 hour integration time \citep{griessmeier11}.

% ###################################################
% Absorption by stellar wind
% ###################################################

\section{Absorption of the planetary radio emission in the stellar wind of the host star}

\begin{figure}
\centering
\includegraphics[width = .4\textwidth]{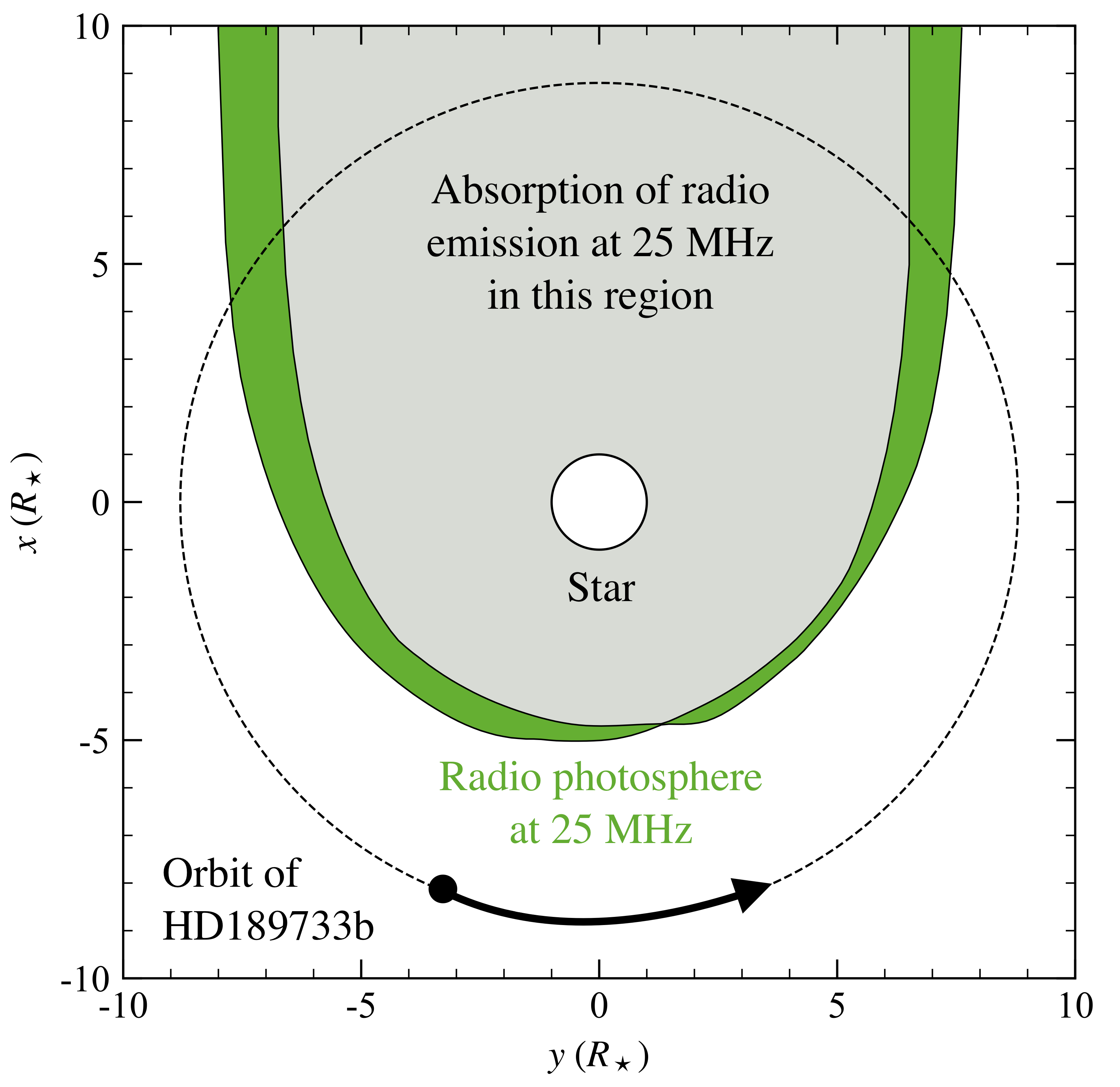}
\includegraphics[width = .4\textwidth]{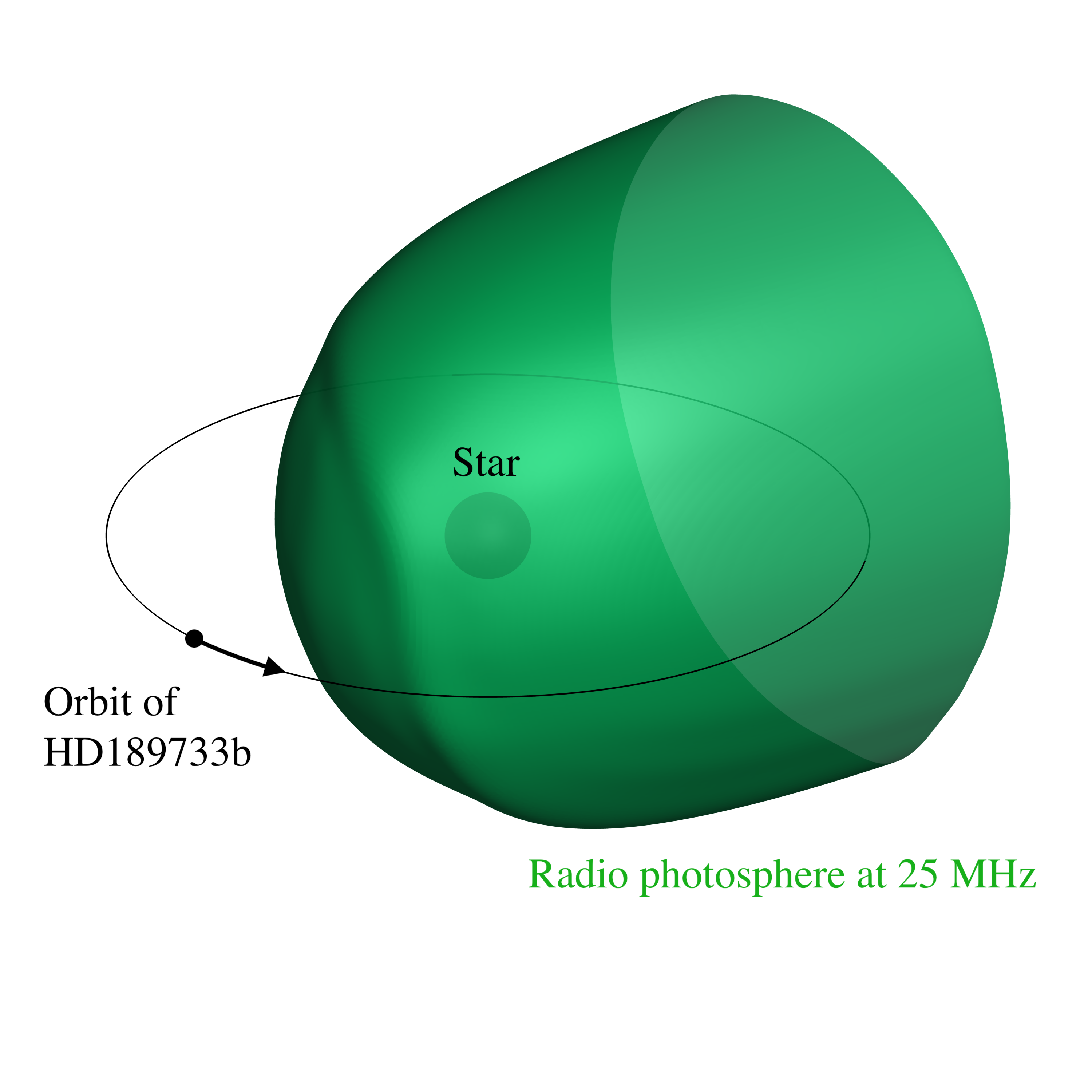}
\caption{The planet orbits through the radio photosphere of the stellar wind, the region optically thick to the emitted planetary frequency of 25~MHz. The left panel shows the shape of the radio photosphere at 25~MHz in the orbital plane, and the right shows its shape in 3D.}
\label{fig:radio photosphere}
\end{figure}

While we predict that radio emission from HD189733b could be detected with LOFAR, the stellar wind itself can absorb low frequency radio emission \citep{panagia75}. Here we solve the equations of radiative transfer for the stellar wind, using the numerical code developed by \citet{ofionnagain19}. We find that the planet orbits through regions of the stellar wind that are optically thick to the predicted frequency emitted from the planet. This is illustrated in Figure~\ref{fig:radio photosphere}. As a result, emission from HD189733b may only be observable as the planet approaches and leave primary transit of the host star. This could be useful information for timing future radio observing campaigns in search of exoplanetary radio emission from systems similar to HD189733b.

% ###################################################
% Conclusions
% ###################################################

\section{Conclusions}

The hot Jupiter HD189733b indeed may be a good target for detecting exoplanetary radio emission. However, as we have shown, the stellar wind can in fact absorb this emission for a large fraction of the planet's orbit. The best time to observe the system is when the planet is near primary transit of the host star. This is also applicable to other exoplanetary systems similar to HD189733b.

% ###################################################
% Acknowledgements
% ###################################################

\section*{Acknowledgements}

RDK acknowledges funding received from the Irish Research Council through the Government of Ireland Postgraduate Scholarship Programme. RDK and AAV also acknowledge funding received from the Irish Research Council Laureate Awards 2017/2018. VB acknowledges support by the Swiss National Science Foundation (SNSF) in the frame of the National Centre for Competence in Research PlanetS, and has received funding from the European Research Council (ERC) under the European Union's Horizon 2020 research and innovation programme (project Four Aces; grant agreement No 724427). This work was carried out using the BATSRUS tools developed at The University of Michigan Center for Space Environment Modeling (CSEM) and made available through the NASA Community Coordinated Modeling Center (CCMC). The authors also wish to acknowledge the SFI/HEA Irish Centre for High-End Computing (ICHEC) for the provision of computational facilities and support.

% ###################################################
% Bibliography
% ###################################################

\def\apj{{ApJ}}    
\def\aap{{A\&A}}   
\def\mnras{{MNRAS}}
\def\aj{{AJ}}
\def\apss{{AP\&SS}}

% ###################################################
% End of document
% ###################################################

\end{document}